\documentclass[runningheads]{svmult}

\usepackage{makeidx}   
\usepackage{graphicx}  
\usepackage{subeqnar}  
\usepackage{multicol}  
\usepackage{physprbb}  
\usepackage{natbib}

\makeindex             

\newcommand{\greeksym}[1]{{\usefont{U}{psy}{m}{n}#1}}
\newcommand{\umu}{\mbox{\greeksym{m}}}

\begin{document}
\title*{GOODS Discovery of a Significant Population of Obscured AGN}
\toctitle{GOODS Discovery of a Significant Population of Obscured AGN}
%
%
\titlerunning{GOODS Discovery of Obscured AGN}
%
\author{C. M. Urry \and Ezequiel Treister}
\authorrunning{C. M. Urry and E. Treister}
%
%
\institute{Yale Center for Astronomy and Astrophysics, 
Yale University, 
P.O. Box 208121, 
New Haven CT 06520-8121, 
USA}

\maketitle              

\begin{abstract}
We analysed the optical and infrared properties of
X-ray sources in the Great Observatories Origins Deep Survey
(GOODS), a deep, multiwavelength survey\index{Surveys,
multiwavelength} covering 0.1 square degrees in two fields.
The HST ACS data\footnote{Based on observations obtained
with the NASA/ESA Hubble Space Telescope, which is operated
by the Association of Universities for Research in Astronomy
(AURA), Inc., under NASA contract NAS5-26555.} are well
explained by a unified AGN scheme that postulates roughly 3
times as many obscured\index{Obscured AGN} as unobscured
AGN, as are the spectroscopic and photometric redshift
distributions once selection effects are considered. Our
model predicts infrared number counts of AGN\index{Active
Galactic Nuclei} that agree well with the preliminary
Spitzer data, confirming that large numbers of obscured AGN
are present in the early Universe ($z>1$).
\end{abstract}

\section{Why There Might be Obscured AGN at High Redshift}
Luminous Active Galactic Nuclei (AGN) have been readily
found throughout the Universe, primarily in surveys at
radio, optical, ultraviolet, and X-ray wavelengths. These
AGN (often called quasars) are ultimately powered by
gravity, as matter from the host galaxy accretes onto a
central supermassive black hole. The prevalence of AGN at
redshifts $z\sim2-3$ indicates that that is the epoch of
greatest black hole growth. It is also an epoch of strong
star formation in young galaxies, one of several arguments
that the growth of supermassive black holes must be closely
tied to the formation of galaxies. Understanding the
co-evolution of galaxies and black holes requires knowing
the accretion history of supermassive black holes, via
an accurate census of black hole demographics:
how many there are, where they are, and what their masses
are.

Although many thousands of AGN have been catalogued, most
recently in the Sloan Digital Sky Survey 
(e.g., \citealp{richards02,anderson03}), the demographics
of supermassive black holes has proved a surprisingly
elusive goal. Very few ``type~2 quasars"\footnote{``Type 2 quasars"
are the luminous analogs of local Seyfert~2 galaxies, i.e., AGN
lacking broad optical emission lines. According to the unification
hypothesis for AGN, the central continuum and broad line region 
in type~2 AGN are obscured by high column densities of gas and dust 
($N_H > 10^{22}$~cm$^{-2}$) along the line of sight.}
have been found
(e.g., \citealp{norman02,stern02,dawson03}). Of course,
UV-excess or optical emission-line surveys would not have
found most obscured AGN, nor would soft X-ray surveys such
as the ROSAT All-Sky \citep{voges99} or WGA \citep{singh95}
surveys. Instead, one needs to look at hard X-rays, where
absorption is a smaller effect, or in the far-infrared, where the
absorbed energy is re-radiated.
 
There are three good reasons to expect many
supermassive black holes to be hidden behind a thick screen
of gas and dust. First, local AGN conform well to the
unification scenario, in which roughly 3/4 of all AGN are at
least partly hidden from a direct line-of-sight
\citep{antonucci93}. Second, the 2--40~keV spectrum of the
X-ray background is very hard, much harder than the typical
spectrum of an unobscured AGN \citep{mushotzky93}, yet high
resolution X-ray imaging makes clear that the ``background"
is in fact summed emission from individual AGN
\citep{giacconi79}. As X-ray astronomers have understood for
nearly two decades, this means most AGN are absorbed by
column densities in excess of $10^{22}$~atoms/cm$^2$
\citep{setti89,madau94,comastri95,gilli01}. Recent papers
(\citealp{comastri03,worsley04,ueda03}; see also Fabian,
this volume) have argued that to explain the $\sim30$~keV
peak in the energy density of the X-ray background requires
a substantial contribution from Compton-thick AGN (with
obscuring column densities $>10^{24}$~cm$^{-2}$). Third,
in the young Universe, where forming galaxies are often
dusty and reddened, obscuration of the central AGN is
even more likely. Thus, at traditional survey wavelengths,
the strong line and nonthermal continuum emission by which
AGN are selected and identified may be completely invisible.

Our strong reliance on optical identification has 
led many to argue that luminous obscured AGN, 
few of which have been found at high redshift, 
do not exist, and the obscuring material must be blown away
in higher luminosity sources. 
While this possibility cannot be ruled out, 
it is certainly premature to explain the absence of obscured AGN 
from surveys that were remarkably insensitive to such objects. 

\section{Searching for Obscured AGN in GOODS}

A strong motivation for the Great Observatories Origins Deep
Survey (GOODS) was the discovery of obscured AGN. GOODS
consists of deep imaging in the far infrared with the
Spitzer Space Telescope \citep{dickinson02} and in the
optical with the Hubble Space Telescope \citep{giavalisco04}
on the footprints of the two deepest Chandra fields
(\citealp{giacconi01,brandt01};\citealp{alexander03}). The
total area is roughly 60 times larger than the original
Hubble Deep Field \citep{williams96} and nearly as deep in
the optical. The Great Observatories data were augmented
with ground-based imaging and
spectroscopy.\footnote{GOODS observations are summarized at
http://www.stsci.edu/science/goods/.} More than 85\% of the
GOODS X-ray sources are AGN, with luminosities $L_X
>10^{42}$~ergs/s; to avoid confusion with starburst
galaxies, we limit our sample to X-ray sources above this
luminosity threshold.

With extensive coverage over five decades in energy from
24~$\umu$m to 8~keV, the GOODS survey is well suited to finding
a high-redshift population of obscured AGN. A
complementary approach, given the relatively low surface
density of AGN compared to normal galaxies, is to target
higher luminosity AGN over a wider area of the sky, an
approach followed by the ChaMP \citep{green04}, CYDER
\citep{castander03}, and HELLAS2XMM \citep{fiore03} surveys,
among others.
Here we describe the AGN detected in the 
X-ray and optical in the GOODS North and South fields,
which represent an order of magnitude more objects 
than in most previous works. 

Most studies of black hole demographics begin with the AGN
content of a given survey, correcting where possible for
selection biases to infer the underlying population. If
selection effects are strong, however, one ends up making
large extrapolations using little information. We therefore
took a different approach: we asked, if there is a
substantial population of obscured AGN, what would be seen
in a deep multiwavelength survey like GOODS? This work is
presented by \citet{treister04}; here we briefly describe
our assumptions and results.

\begin{figure}[t]
\begin{center}
\includegraphics[width=.43\textwidth]{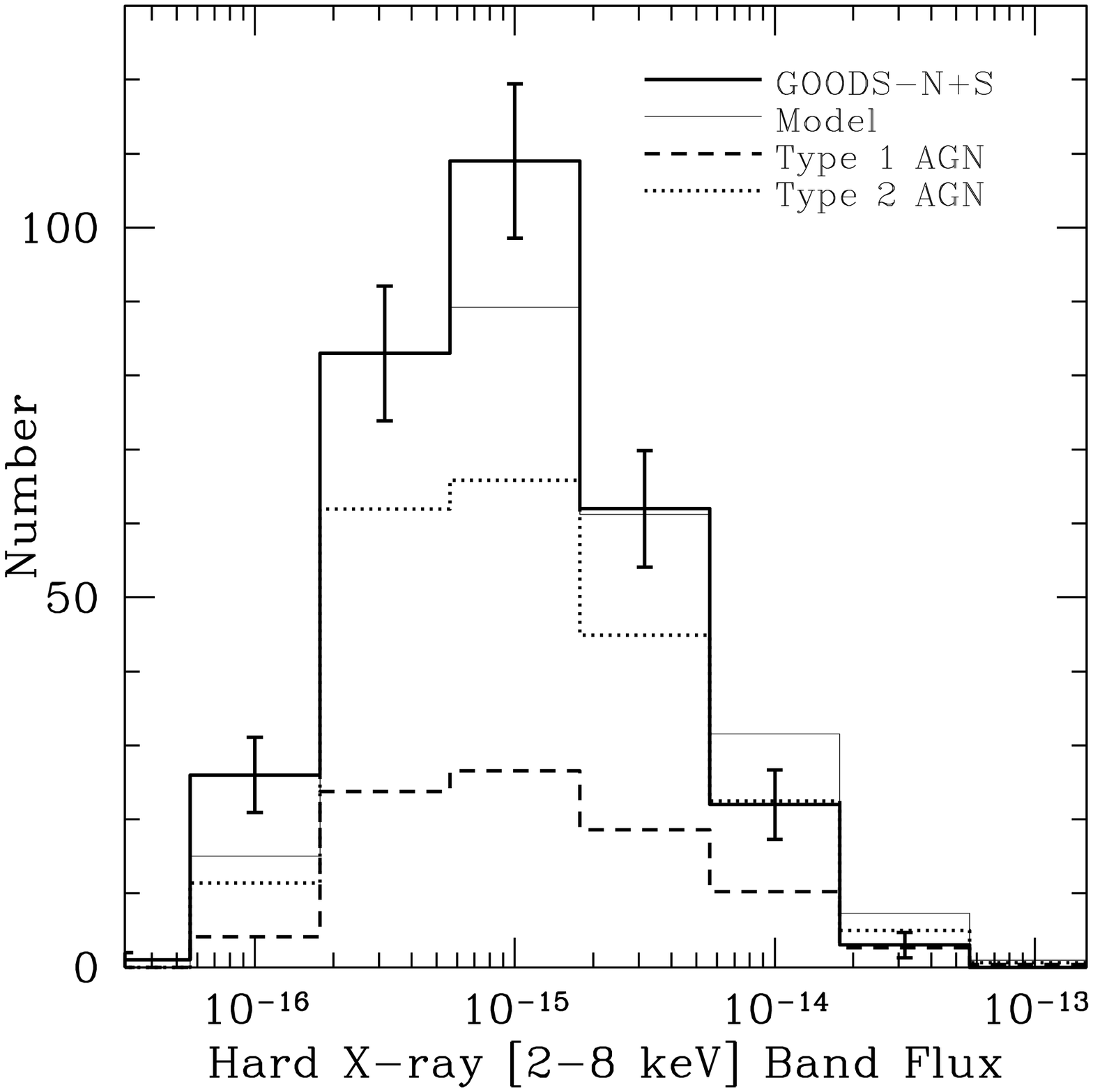}
\includegraphics[width=.43\textwidth]{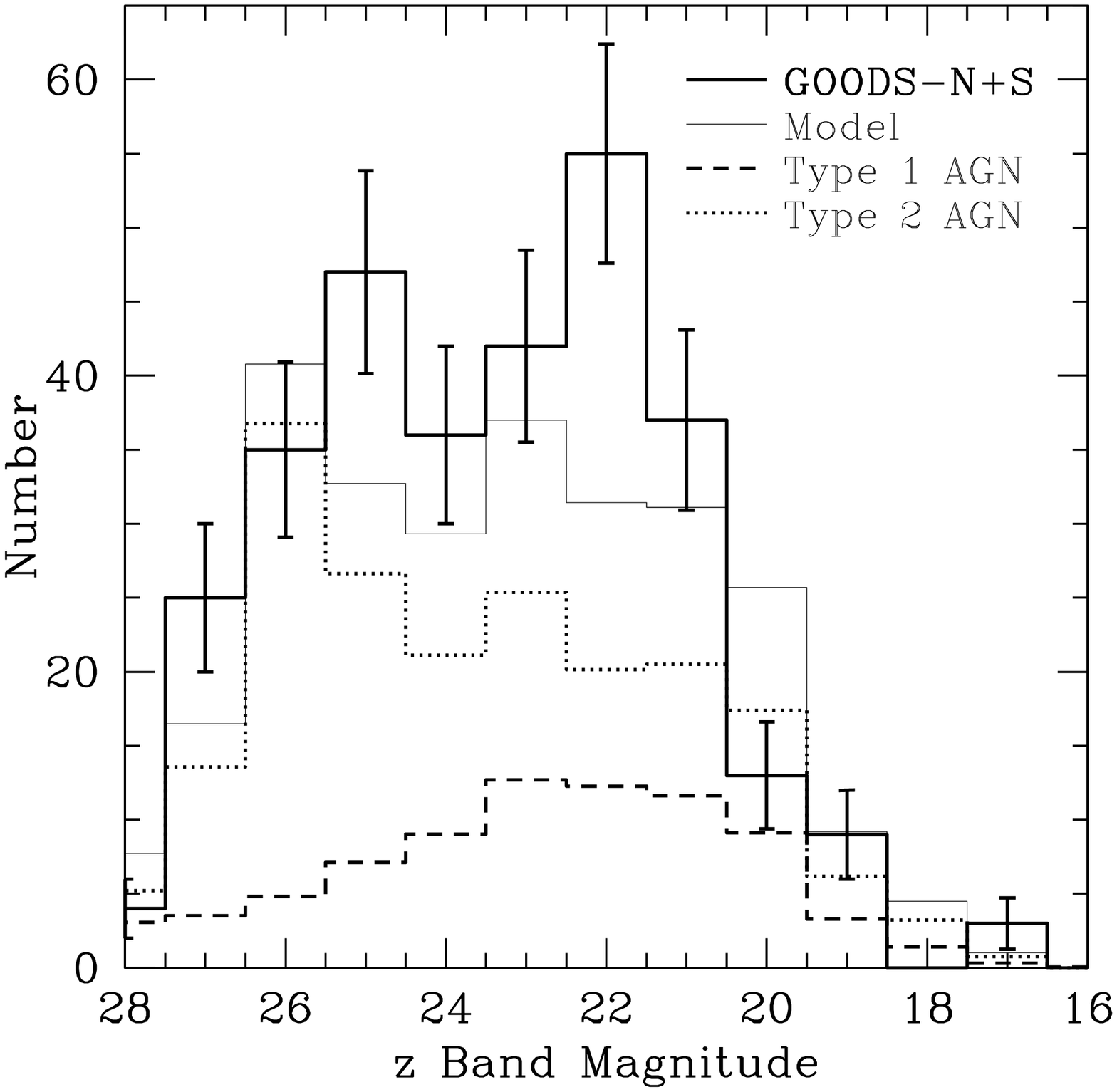}
\caption[]{X-ray fluxes and z-band magnitudes of GOODS AGN. 
{\it (Left)} Hard X-ray flux distribution
for X-ray sources in the combined
GOODS-North and -South fields ({\it heavy solid line}),
compared to number counts
calculated from a simple unification model
({\it light solid line}). 
Contributions from unobscured ({\it dashed line}) and 
obscured ({\it dotted line}) AGN are shown separately;
these follow similar distributions 
because hard X-rays are not strongly affected by
absorption.
{\it (Right)} Distribution of observed $z$-band magnitudes 
for GOODS-North and -South X-ray sources
({\it heavy solid line}), compared to model distribution 
({\it light solid line}). 
Obscured AGN ({\it dotted line}) dominate at faint magnitudes,
where only the host galaxy contributes to the optical light,
while bright samples consist both unobscured 
({\it dashed line}) and obscured AGN. 
}
\label{fig:F_X_opt}
\end{center}
\end{figure}

We start with a very simple unification scheme for AGN, in
which a central engine -- well understood from studies of
unobscured AGN -- is surrounded by a dusty
torus\footnote{Note that the geometry need not be a torus; 
we use this geometry simply to derive the distribution
of absorbing column densities expected from random
orientations. The resulting $N_H$ distribution matches well
the observed one, with the proviso that Compton-thick
sources are missing, but so would many other geometries.} 
that absorbs optical through soft X
radiation and re-radiates the absorbed energy as thermal
infrared emission. We combine well-measured spectra of
unobscured AGN with plausible reddening and absorption laws,
and with the dust emission models of \citet{elitzur03} and
\citet{nenkova02}, to create a grid of spectra appropriate
to increasing absorbing column densities along the line of
sight to the AGN. The torus geometry was adjusted 
so that three-quarters of all AGN are obscured 
(i.e., have $N_H > 10^{22}$~cm$^{-2}$) and so the 
mid-plane column density is $10^{24}$~cm$^{-2}$ (even the
Chandra deep surveys are not sensitive to objects with
higher column densities).
We further assumed this geometry was independent of redshift
and luminosity -- the simplest assumption, and one that,
frankly, we expected to prove wrong. We added an $L_*$ host
galaxy spectrum to each AGN template spectrum. Finally, we
assumed the most recent, hard-X-ray-selected luminosity
function and luminosity-dependent density evolution given by
\citet{ueda03}.

As shown in Figure~1, this very simple model explains very
well the X-ray and optical counts in the GOODS fields.
That it explains the X-ray counts is not too surprising since
we start with a hard X-ray luminosity function and evolution
derived from these fields, among others. 
For the optical, however, the agreement indicates that
faint optical sources are primarily obscured AGN.
Indeed, these tend to have red colors and hard X-ray spectra,
as expected of obscured AGN.

These results mean that the GOODS HST data agree well 
with a population of obscured AGN extending from $z=0$ 
to well beyond $z>1$.
Specifically, the X-ray and optical properties of GOODS 
AGN are fully consistent with
a very simple unification picture, 
requiring no modification with
either redshift or AGN luminosity. 
The implication is that some X-ray-emitting AGN are missed
in the Chandra Deep Field samples, and others lack optical
spectroscopic identifications. Since the best 8- or 10-m
class telescopes have an effective limit at $R<24$~mag,
many obscured AGN will lack spectroscopic identification.

The redshift distribution is a critical point. The best
population synthesis models for the X-ray background prior
to the new Chandra and XMM deep fields \citep{gilli01}
incorporated a high-redshift population of obscured AGN that
evolves similarly to the unobscured AGN, and hence predicted
a redshift distribution peaking at $z\sim1.5$. Yet observed
redshift distributions from the Chandra Deep Fields and the
Lockman hole peak at much lower redshifts (e.g.,
\citealp{hasinger02,barger03}). This has been taken as
evidence of a fatal flaw in the population synthesis models.

We find the discrepancy does not imply any lack of obscured
quasars at high redshift. Rather, it results from a
combination of two factors: (1) the luminosity-dependent
density evolution now measured for X-ray AGN, which was not
used in the \citet{gilli01} models and which increases the
fraction of AGN at lower luminosity and lower redshift, and
more importantly, (2) the implicit spectroscopic limit for X-ray samples,
which preferentially excludes the faint counterparts of
high-redshift, obscured AGN.

The latter point is illustrated in Figure~2, 
which shows the redshift distribution for GOODS AGN.
The upper dashed line shows the predicted redshift
distribution for our unified model, which still peaks above
a redshift of 1, much like the earlier population synthesis models.
If we impose a spectroscopic limit of $R=24$~mag, 
however, we obtain the lower dashed line, which agrees
well with the observed spectroscopic redshift distribution. 
Note that the photometric redshifts 
are indeed shifted to higher redshifts compared to
the spectroscopic redshifts, although they too miss
the very faintest, highest redshift obscured AGN.

\begin{figure}[t]
\begin{center}
\includegraphics[width=.45\textwidth]{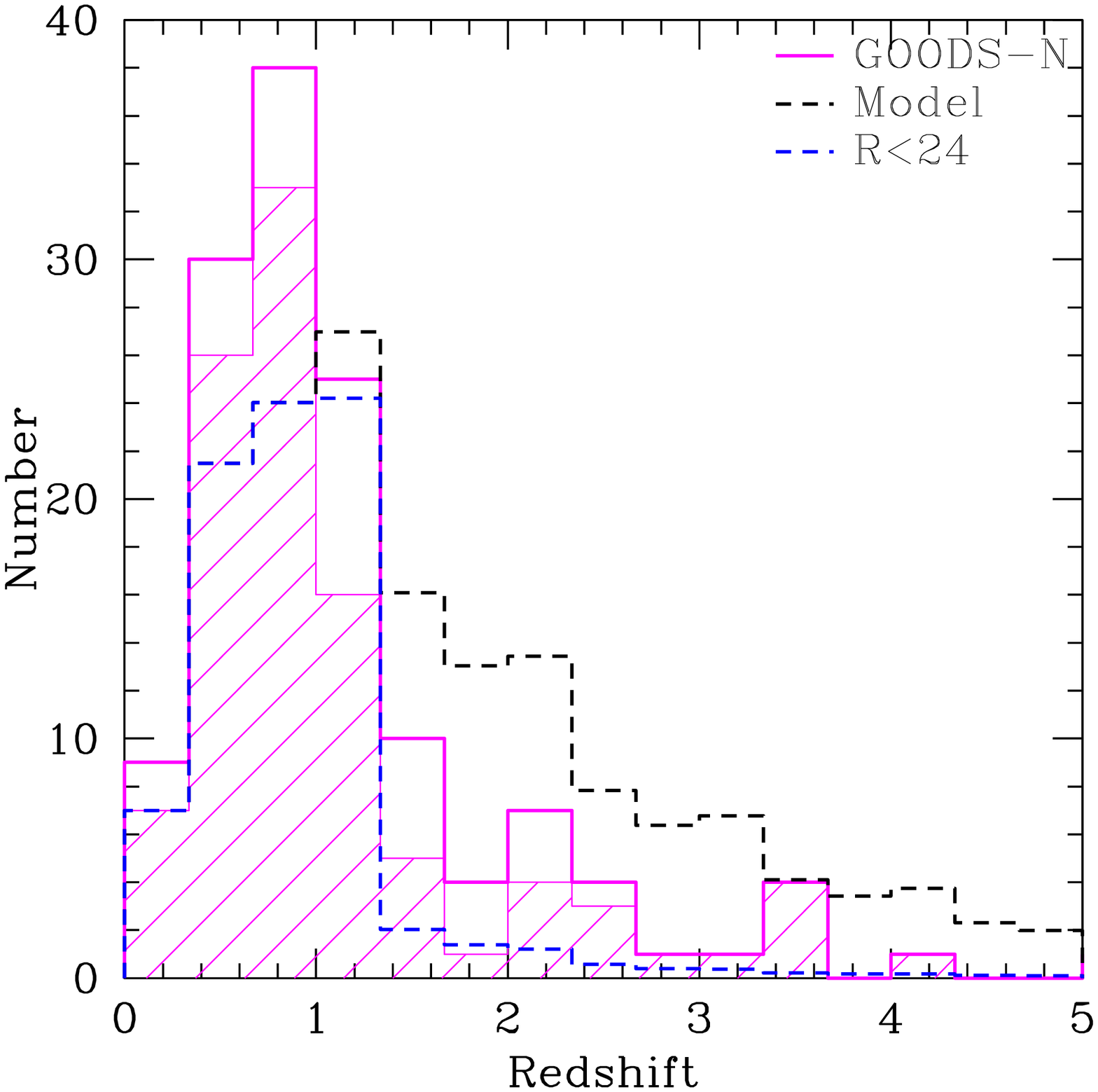}
\includegraphics[width=.45\textwidth]{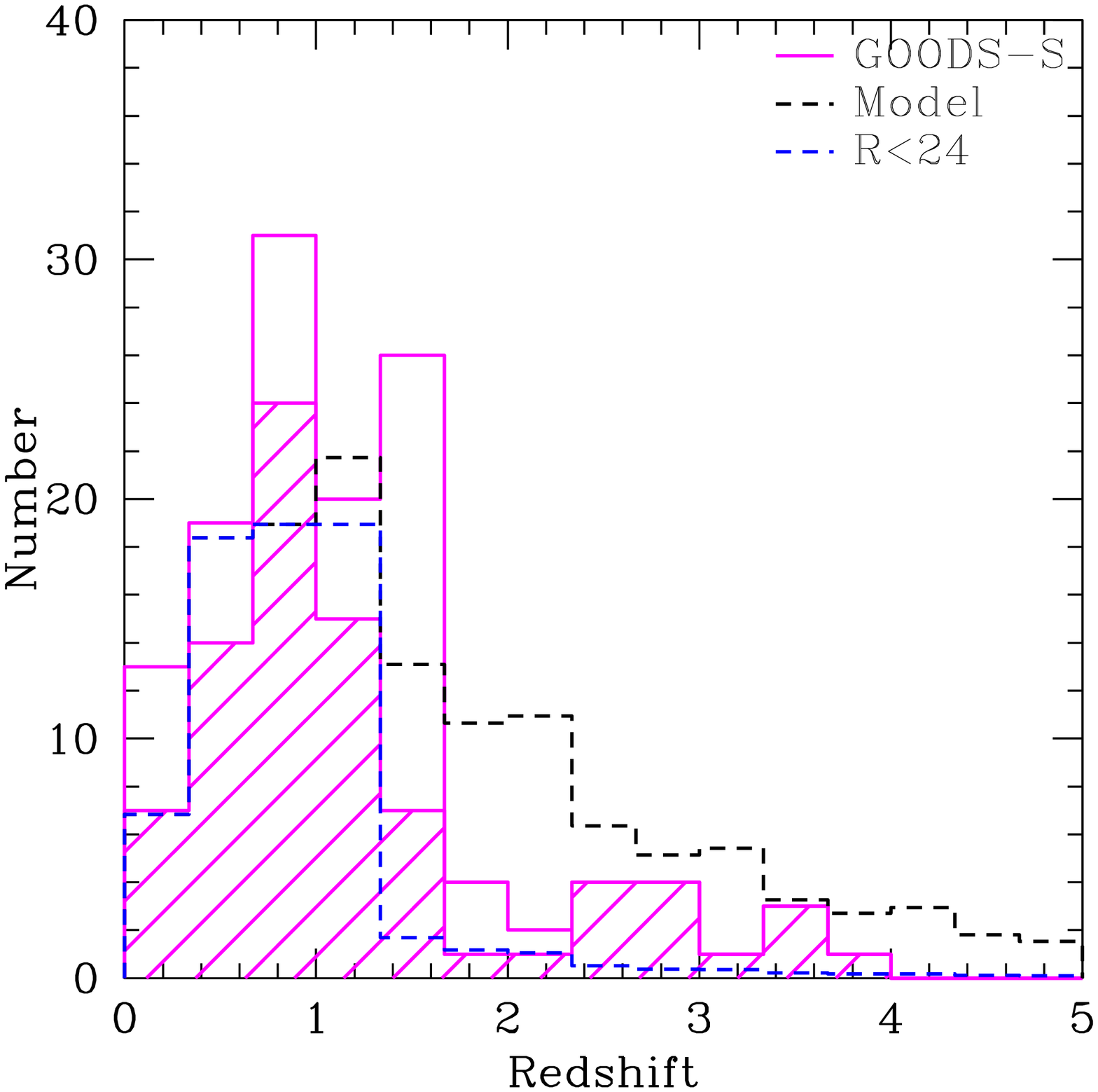}
\end{center}
\caption[]{Observed redshift distributions 
{\it (heavy solid line)} 
for AGN in the GOODS-South {\it (left panel)} and -North
{\it (right panel)} fields, including both spectroscopic
{\it (hatched area}; \citealp{barger03,szokoly04}) and
photometric redshifts (\citealp{barger03,mobasher04}); these are
100\% complete in the GOODS-S field and 75\% complete in the
GOODS-N region. The predicted distribution ({\it upper dashed
line}) peaks above a redshift of 1, similar to earlier
population synthesis models and in apparent disagreement
with the observed one. However, if we cut the sample at
$R<24$~mag ({\it lower dashed line}), as appropriate for a
sample with optical spectra, the model and observed
distributions agree well. The data are
therefore consistent with a significant high-redshift
population of obscured AGN which are missed in spectroscopic
samples due to their faint optical magnitudes.
}
\label{fig:zdis}
\end{figure}

\section{Spitzer Data and the Far-Infrared Properties of AGN}

According to the unification scheme, 
dust and gas absorbs much of the luminous optical 
through X-ray AGN emission, and is thus heated.
The absorbed energy is re-radiated as thermal 
emission, which should be readily visible in the
far-infrared. 
A key element of the GOODS project is very deep imaging
with the Spitzer observatory, in the IRAC bands from
3.6 to 8.4 microns and the MIPS 24-micron band.
These data will provide a strong test of our unified model.

We used our simple unification model to predict the
far-infrared source counts, in each Spitzer band, for the
appropriate GOODS flux limits \citep{treister04}. As of
spring 2004, some IRAC data in the GOODS-S field were
already in hand. We compared preliminary photometry of the
AGN to our source count prediction. The results are shown
in Figure~3 (van Duyne et~al., in prep.), 
along with predictions for the 24-micron source
counts. (The first GOODS MIPS data will be obtained in
summer/fall 2004.)

\begin{figure}[t]
\begin{center}
\includegraphics[width=.55\textwidth]{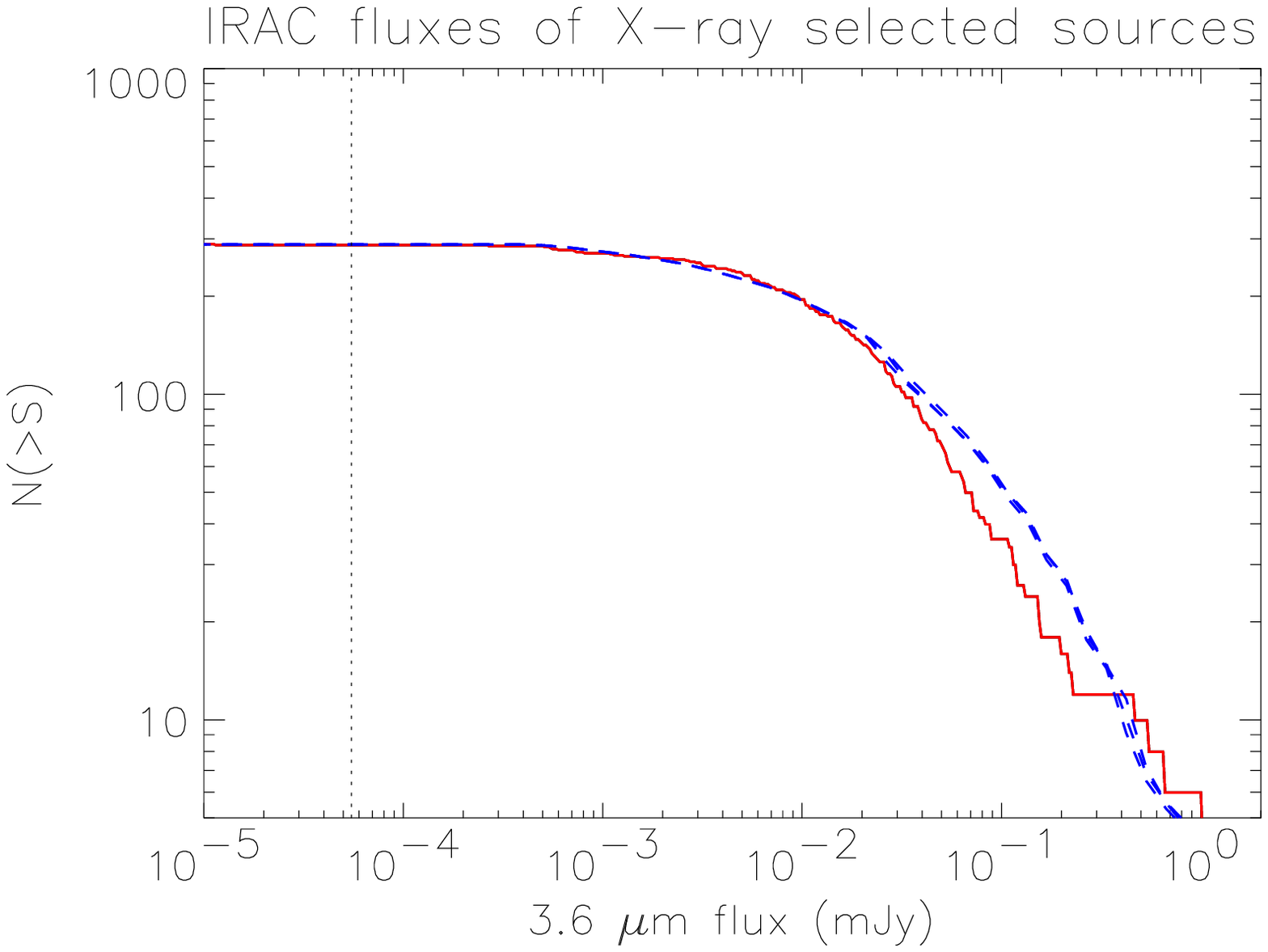}
\includegraphics[width=.38\textwidth]{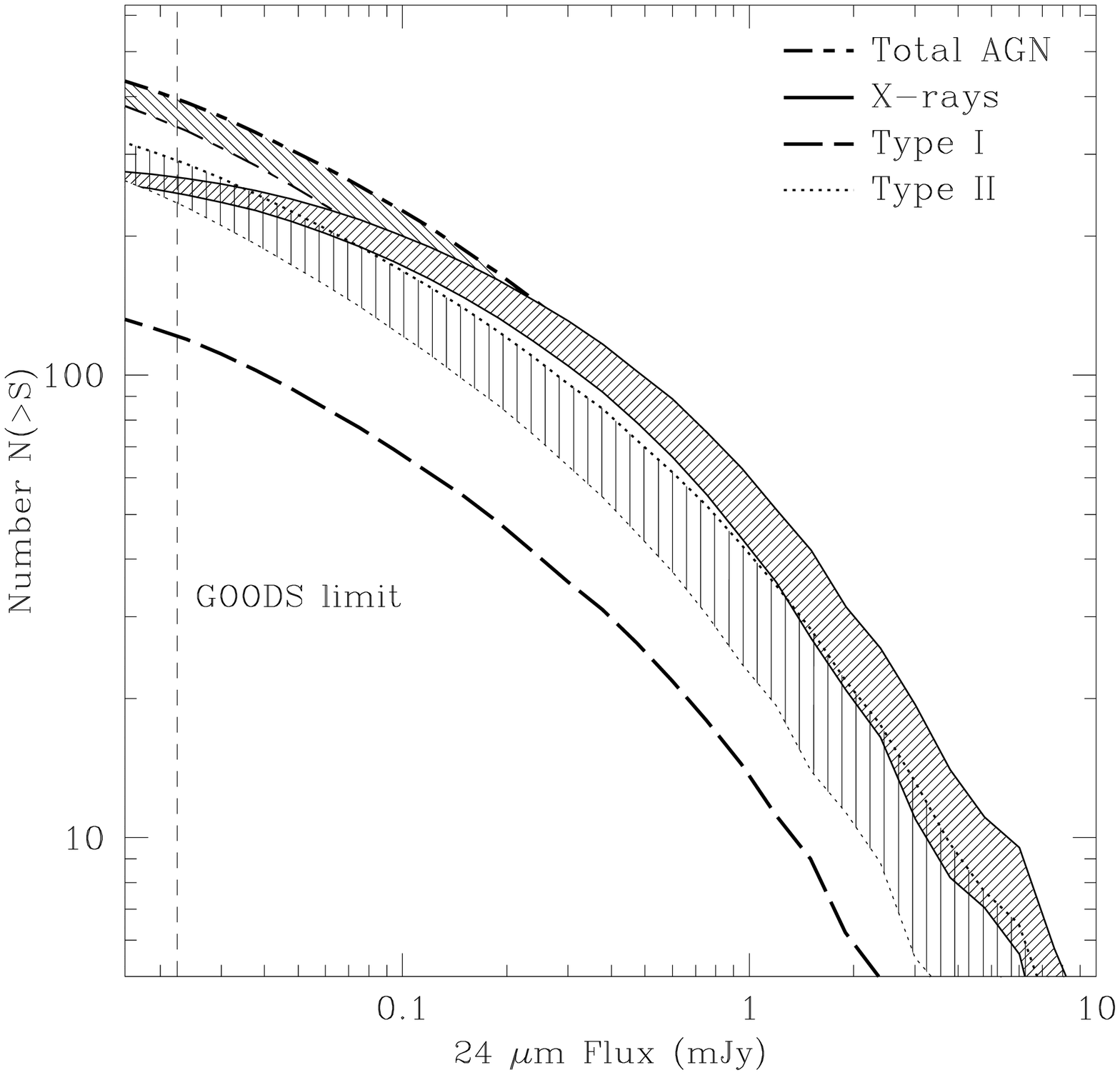}
\end{center}
\caption[]{Predicted Spitzer counts for the simple unified model.
{\it (Left)} 
The model prediction at 3.6-$\umu$m {\it (dashed line)}
agrees well with the preliminary Spitzer counts of
GOODS-S X-ray sources {\it (solid histogram)}, 
in both normalization and shape.
{\it (Right)} 
Model prediction at 24~$\umu$m for total source counts,
({\it upper dot-dash line}) and for
AGN brighter than the X-ray flux limit of GOODS
({\it solid line});
the difference represents the AGN missed by Chandra. 
Dotted and dashed lines represent the obscured and
unobscured AGN, respectively; the former dominate by
the specified factor 3:1.}
\label{fig:IRcounts}
\end{figure}

Certainly these early data require closer inspection,
and the model awaits a stronger test from
the full multiwavelength Spitzer data set.
The pending data will allow us to verify whether large numbers 
of obscured AGN are indeed present in the early Universe,
and will also be extremely important for
refining the AGN model, specifically
the covering factor,
dust geometry, and evolution thereof.
The 24-micron images are particularly powerful,
since the AGN emission is most isotropic
at the longest infrared wavelengths.
Obscured AGN should be very bright far-infrared sources;
those that are missed even by X-ray observations
probably look like ultraluminous infrared galaxies.

We have looked at the spectral energy distributions
of the GOODS AGN, including the available Spitzer data.
Indeed, the hard X-ray sources with faint, red optical
counterparts are extra-luminous in the Spitzer bands.
Two example spectral energy distributions are shown in Figure~4
(van Duyne et~al., in prep.).
Both cases look just like a normal galaxies at optical
wavelengths, but are revealed to be AGN by the
luminous X-ray emission.
They are clearly obscured by virtue of their
faint optical magnitudes, hard X-ray spectra,
and bright infrared fluxes.

\begin{figure}
\begin{center}
\includegraphics[width=.45\textwidth]{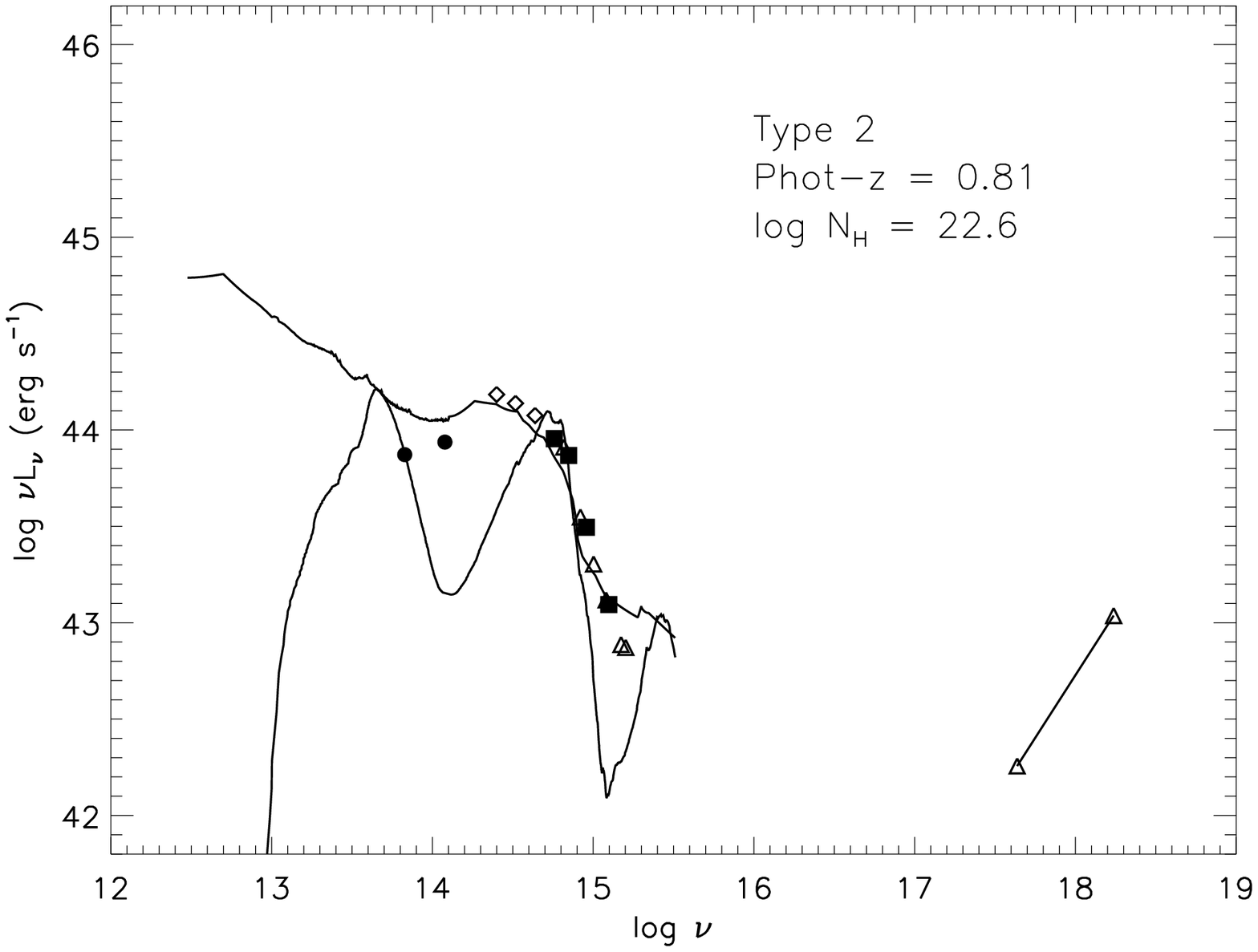}
\includegraphics[width=.45\textwidth]{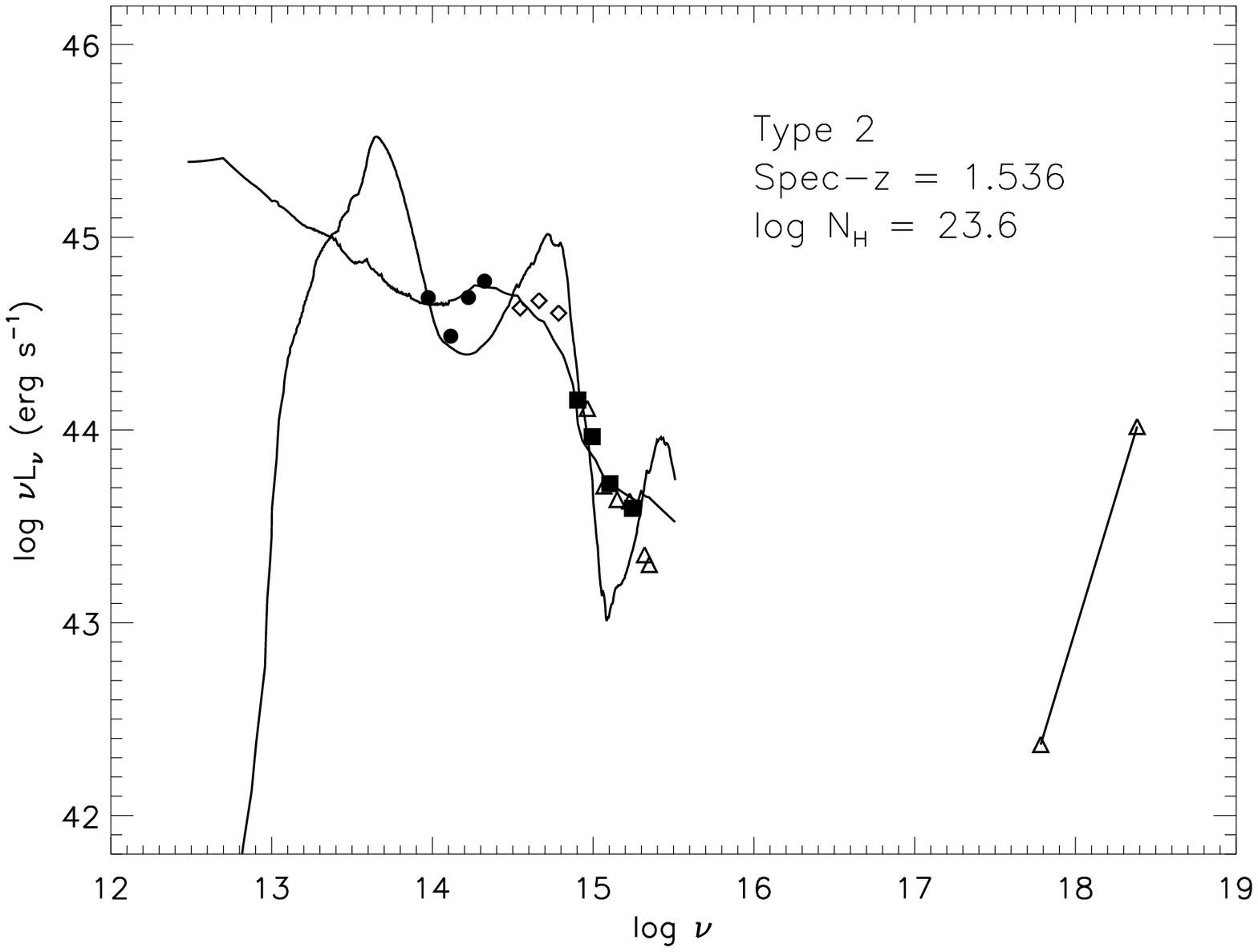}
\end{center}
\caption[]{Spectral energy distributions of two GOODS AGN,
including HST ACS data {\it (squares)}, Chandra X-ray data
{\it (triangles, straight lines)}, and Spitzer data 
{\it (filled circles)}.
Also shown are the best-fit Seyfert~2 galaxy templates,
both observed {\it (light solid line)} and model spectra
{\it (heavy solid line)}.
} 
\label{fig:SEDs}
\end{figure}

\section{Summary and Conclusions}

The GOODS multiwavelength survey supports the idea of a
significant population of obscured AGN persisting to high
redshift. From a strictly observational viewpoint, we
cannot rule out this population, and indeed, see every
evidence for its existence. In a subsequent paper \citep{treister04b}
we explain the
apparent decline in the ratio of obscured to unobscured AGN
with redshift or luminosity, which is seen in several
X-ray-selected AGN samples (\citealp{hasinger03}; U03), 
by the same selection effects against Compton thick and/or optically
faint sources.

The data are consistent with a simple unified model of a
luminous AGN nucleus obscured along some lines of sight
by dust and gas. Our work is not strongly sensitive to the
geometry, and other configurations of dust are equally
possible as long as they predict a similar distribution of
absorbing column densities. We are also not sensitive to
the numbers of Compton-thick AGN,
a population that will be probed
with future X-ray missions sensitive in the energy range
10-100~keV. In any case, we conclude that at least 3/4 of
all supermassive black holes are obscured by column
densities of at least $N_H \sim 10^{22}$~cm$^{-2}$.

The unified model implies that half the obscured AGN are not
detected in X-ray fields at the Chandra Deep Field depths.
They should be clearly visible in the Spitzer data, although
separating heavily obscured AGN from very luminous starbursts
will not be simple. Upcoming Spitzer data, along with the
integral constraint from the X-ray ``background" spectrum,
will better constrain the geometry and other properties of
the obscuring material.

Individual spectral energy distributions of the AGN will
also constrain the dust geometry, and perhaps the
contribution from any starburst component. Bolometric
corrections in the far-infrared are likely to be
significant, and affect estimates of the mass accretion rate
and thus the growth rate of supermassive black holes. Note
that observed spectral energy distributions 
never represent the actual output from the AGN engine: the
luminosity of unobscured AGN is overestimated
(since the optical through soft X-ray emission is not
isotropic), while obscured AGN appear
under-luminous relative to their intrinsic power. 
Bolometric
corrections can be calculated from models, such as those
considered here, but can not be derived empirically for any
highly anisotropic source. This is critical for
understanding the energetics and structures of AGN.

One strong conclusion from our work is that one should look
for obscured black holes where they can be easily found,
namely in the far-infrared and hard X-ray. Any substantial
and widespread obscuration has important implications for
the cosmic accretion history of black holes, and for the
average efficiency of converting gravitational potential
energy to radiation. 

{\it Acknowledgements} -- This work, which was supported in part
by NASA grant HST-GO-09425.13-A,
owes a great deal to our colleagues
in the GOODS project, particularly Yale graduate student
Jeff van Duyne and the co-authors of the
\citet{treister04} paper.

%


\begin{thebibliography}{}

\bibitem[{{Alexander} {et~al.} (2003) {Alexander}, {Bauer}, {Brandt},
  {Schneider}, {Hornschemeier}, {Vignali}, {Barger}, {Broos}, {Cowie},
  {Garmire}, {Townsley}, {Bautz}, {Chartas}, \& {Sargent}}]{alexander03}
{Alexander}, D.~M., {Bauer}, F.~E., {Brandt}, W.~N., {Schneider}, D.~P.,
  {Hornschemeier}, A.~E., {Vignali}, C., {Barger}, A.~J., {Broos}, P.~S.,
  {Cowie}, L.~L., {Garmire}, G.~P., {Townsley}, L.~K., {Bautz}, M.~W.,
  {Chartas}, G., \& {Sargent}, W.~L.~W. 2003, AJ, 126, 539

\bibitem[Anderson et~al.(2003)]{anderson03} Anderson, S. F., et~al.\ 2003,
AJ, 126, 2209

\bibitem[{{Antonucci}(1993)}]{antonucci93}{Antonucci}, R. 1993, ARA\&A, 31, 473

\bibitem[{{Barger} {et~al.}(2003){Barger}, {Cowie}, {Capak}, {Alexander},
  {Bauer}, {Fernandez}, {Brandt}, {Garmire}, \& {Hornschemeier}}]{barger03}
{Barger}, A.~J., {Cowie}, L.~L., {Capak}, P., {Alexander}, D.~M., {Bauer},
  F.~E., {Fernandez}, E., {Brandt}, W.~N., {Garmire}, G.~P., \&
  {Hornschemeier}, A.~E. 2003, AJ, 126, 632

\bibitem[{{Brandt} {et~al.}(2001) {Brandt}, {Alexander}, {Hornschemeier},
  {Garmire}, {Schneider}, {Barger}, {Bauer}, {Broos}, {Cowie}, {Townsley},
  {Burrows}, {Chartas}, {Feigelson}, {Griffiths}, {Nousek}, \&
  {Sargent}}]{brandt01}
{Brandt}, W.~N., {Alexander}, D.~M., {Hornschemeier}, A.~E., {Garmire}, G.~P.,
  {Schneider}, D.~P., {Barger}, A.~J., {Bauer}, F.~E., {Broos}, P.~S., {Cowie},
  L.~L., {Townsley}, L.~K., {Burrows}, D.~N., {Chartas}, G., {Feigelson},
  E.~D., {Griffiths}, R.~E., {Nousek}, J.~A., \& {Sargent}, W.~L.~W. 2001, AJ,
  122, 2810

\bibitem[Castander et al.(2003)]{castander03} Castander, F.~J., 
Treister, E., Maza, J., Coppi, P.~S., Maccarone, T.~J., Zepf, S.~E., 
Guzm{\' a}n, R., \& Ruiz, M.~T.\ 2003, Astronomische Nachrichten, 324, 40 

\bibitem[{{Comastri} {et~al.}(1995){Comastri}, {Setti}, {Zamorani}, {Hasinger}}]{comastri95}
{Comastri}, A., {Setti}, G., {Zamorani}, G. \& {Hasinger}, G. 1995, A\&A, 296, 1

\bibitem[{{Comastri} {et~al.}(2003){Comastri} and others}]{comastri03} 
Comastri, A.~\& the HELLAS2XMM team 2003, Societa Astronomica Italiana 
Memorie Supplement, 3, 179 

\bibitem[{{Dawson} {et~al.}(2003){Dawson}, {McCrady}, {Stern}, {Eckart},
  {Spinrad}, {Liu}, \& {Graham}}]{dawson03}
{Dawson}, S., {McCrady}, N., {Stern}, D., {Eckart}, M.~E., {Spinrad}, H.,
  {Liu}, M.~C., \& {Graham}, J.~R. 2003, AJ, 125, 1236

\bibitem[{{Dickinson} \& {Giavalisco}(2002)}]{dickinson02}
{Dickinson}, M. \& {Giavalisco}, M. 2002, in The Mass of Galaxies at Low and
  High Redshift, in press, astro-ph/0204213

\bibitem[{{Elitzur} {et~al.}(2003){Elitzur}, {Nenkova}, \&
  {Ivezic}}]{elitzur03}
{Elitzur}, M., {Nenkova}, M., \& {Ivezic}, Z. 2003, {astro-ph/0309040}

\bibitem[Fiore et al.(2003)]{fiore03} Fiore, F., et al.\ 2003, 
A\&A, 409, 79 

\bibitem[{{Giacconi} {et~al.}(1979){Giacconi}, {Bechtold}, {Branduardi},
  {Forman}, {Henry}, {Jones}, {Kellogg}, {van der Laan}, {Liller}, {Marshall},
  {Murray}, {Pye}, {Schreier}, {Sargent}, {Seward}, \&
  {Tananbaum}}]{giacconi79}
{Giacconi}, R., {Bechtold}, J., {Branduardi}, G., {Forman}, W., {Henry}, J.~P.,
  {Jones}, C., {Kellogg}, E., {van der Laan}, H., {Liller}, W., {Marshall}, H.,
  {Murray}, S.~S., {Pye}, J., {Schreier}, E., {Sargent}, W.~L.~W., {Seward},
  F., \& {Tananbaum}, H. 1979, ApJ, 234, L1

\bibitem[{{Giacconi} {et~al.}(2001){Giacconi}, {Rosati}, {Tozzi}, {Nonino},
  {Hasinger}, {Norman}, {Bergeron}, {Borgani}, {Gilli}, {Gilmozzi}, \&
  {Zheng}}]{giacconi01}
{Giacconi}, R., {Rosati}, P., {Tozzi}, P., {Nonino}, M., {Hasinger}, G.,
  {Norman}, C., {Bergeron}, J., {Borgani}, S., {Gilli}, R., {Gilmozzi}, R., \&
  {Zheng}, W. 2001, ApJ, 551, 624

\bibitem[{{Giavalisco} {et~al.} (2004)}]{giavalisco04}
{Giavalisco}, M. {et~al.} 2004, ApJ, 600, L93

\bibitem[{{Gilli} {et~al.}(2001){Gilli}, {Salvati}, \& {Hasinger}}]{gilli01}
{Gilli}, R., {Salvati}, M., \& {Hasinger}, G. 2001, A\&A, 366, 407

\bibitem[Green et al.(2004)]{green04} Green, P.~J., et al.\ 
2004, ApJS, 150, 43 

\bibitem[{{Hasinger}(2002)}]{hasinger02}
{Hasinger}, G. 2002, in New Visions of the X-ray Universe in the XMM-Newton and Chandra Era, 
Ed. F. Jansen (ESTEC: ESA SP-488), XX (astro-ph/0202430)

\bibitem[{Hasinger(2004)}]{hasinger03}
Hasinger, G. 2004, Nucl. Phys. Proc. Suppl., 132, 86

\bibitem[Madau, Ghisellini, \& Fabian(1994)]{madau94} Madau, 
P., Ghisellini, G., \& Fabian, A.~C.\ 1994, MNRAS, 270, L17 

\bibitem[{{Mobasher} {et~al.}(2004)}]{mobasher04}
{Mobasher}, B. {et~al.} 2004, ApJ, 600, L167

\bibitem[{{Mushotzky} {et~al.}(1993){Mushotzky}, {Done}, \&
  {Pounds}}]{mushotzky93} {Mushotzky}, R.~F., {Done}, C., \&
  {Pounds}, K.~A. 1993, ARA\&A, 31, 717


\bibitem[{{Nenkova} {et~al.}(2002){Nenkova}, {Ivezi{\' c}}, \&
  {Elitzur}}]{nenkova02}
{Nenkova}, M., {Ivezi{\' c}}, {\v Z}., \& {Elitzur}, M. 2002, ApJ, 570, L9

\bibitem[{{Norman} {et~al.}(2002){Norman}, {Hasinger}, {Giacconi}, {Gilli},
  {Kewley}, {Nonino}, {Rosati}, {Szokoly}, {Tozzi}, {Wang}, {Zheng}, {Zirm},
  {Bergeron}, {Gilmozzi}, {Grogin}, {Koekemoer}, \& {Schreier}}]{norman02}
{Norman}, C., {Hasinger}, G., {Giacconi}, R., {Gilli}, R., {Kewley}, L.,
  {Nonino}, M., {Rosati}, P., {Szokoly}, G., {Tozzi}, P., {Wang}, J., {Zheng},
  W., {Zirm}, A., {Bergeron}, J., {Gilmozzi}, R., {Grogin}, N., {Koekemoer},
  A., \& {Schreier}, E. 2002, ApJ, 571, 218

\bibitem[Richards et~al.(2002)]{richards02} Richards, G. T., et~al.\ 2002,
AJ, 123, 2945

\bibitem[{{Setti} \& {Woltjer}(1989){Setti}, {Woltjer}}]{setti89}
{Setti}, G., {Woltjer}, L. 1989, A\&A, 224L, 21

\bibitem[{{Singh} {et~al.}(1995){Singh}, {Barrett}, {White}, {Giommi}, \&
  {Angelini}}]{singh95}
{Singh}, K.~P., {Barrett}, P., {White}, N.~E., {Giommi}, P., \& {Angelini}, L.
  1995, ApJ, 455, 456


\bibitem[{{Stern} {et~al.}(2002){Stern}, {Moran}, {Coil}, {Connolly}, {Davis},
  {Dawson}, {Dey}, {Eisenhardt}, {Elston}, {Graham}, {Harrison}, {Helfand},
  {Holden}, {Mao}, {Rosati}, {Spinrad}, {Stanford}, {Tozzi}, \& {Wu}}]{stern02}
{Stern}, D., {Moran}, E.~C., {Coil}, A.~L., {Connolly}, A., {Davis}, M.,
  {Dawson}, S., {Dey}, A., {Eisenhardt}, P., {Elston}, R., {Graham}, J.~R.,
  {Harrison}, F., {Helfand}, D.~J., {Holden}, B., {Mao}, P., {Rosati}, P.,
  {Spinrad}, H., {Stanford}, S.~A., {Tozzi}, P., \& {Wu}, K.~L. 2002, ApJ,
  568, 71

\bibitem[{{Szokoly} {et~al.}(2004)}]{szokoly04}
{Szokoly}, G. {et~al.} 2004, ApJS {in press, astro-ph/0312324}


\bibitem[{{Treister} {et~al.}(2004a){Treister} and others}]{treister04} 
{Treister}, E. {et al.} 2004a, ApJ in press, arXiv:astro-ph/0408099.

\bibitem[{{Treister} {et~al.}(2004b){Treister} and others}]{treister04b}
{Treister}, E. {et al.} 2004b, ApJ submitted.

\bibitem[{{Ueda} {et~al.}(2003; U03){Ueda}, {Akiyama}, {Ohta}, \& {Miyaji}}]{ueda03}
{Ueda}, Y., {Akiyama}, M., {Ohta}, K., \& {Miyaji}, T. 2003,
ApJ, 598, 886 (U03)


\bibitem[{{Voges} {et~al.}(1999)}]{voges99}
{Voges}, W. {et~al.} 1999, A\&A, 349, 389

\bibitem[Williams et al.(1996)]{williams96} Williams, R.~E., et 
al.\ 1996, AJ, 112, 1335 

\bibitem[Worsley et al.(2004)]{worsley04} Worsley, M.~A., Fabian, 
A.~C., Barcons, X., Mateos, S., Hasinger, G., \& Brunner,
H.\ 2004, MNRAS, 352, L28

\end{thebibliography}
\end{document}